*Research Article*

# Anisotropic Quintessence Strange Stars in $f(T)$ Gravity with Modified Chaplygin Gas

**Pameli Saha  and Ujjal Debnath**

*Department of Mathematics, Indian Institute of Engineering Science and Technology, Shibpur, Howrah 711 103, India*

Correspondence should be addressed to Pameli Saha; pameli.saha15@gmail.com

In this paper, we study the existence of strange star in the background of $f(T)$ modified gravity where $T$ is a scalar torsion. In KB metric space, we derive the equations of motion using anisotropic property within the spherically strange star with modified Chaplygin gas in the framework of modified $f(T)$ gravity. Then we obtain many physical quantities to describe the physical status such as anisotropic behavior, energy conditions, and stability. By the matching condition, we calculate the unknown parameters to evaluate the numerical values of mass, surface redshift, etc., from our model to make comparison with the observational data.

## 1. Introduction

In modern cosmology, cosmic acceleration is an interesting discovery. The observation of type Ia supernovae (SNeIa) together with the cosmic microwave background (CMB), large scale structure surveys (LSS), and Wilkinson Microwave Anisotropy Probe (WMAP) [1–4] ensures the presence of an exotic energy component dominating our universe which is entitled as dark energy (DE) **having equation of state $p = w\rho$ with strong negative pressure. For accelerating expansion $w$ must satisfy the range $w < -1/3$. If $-1 < w < -1/3$ then it belongs to quintessence phase and if $w < -1$, then it belongs to phantom regime. In particular, when $w = -1 \implies p = -\rho$ then the equation of state of the interior region of a Gravastar (gravitationally vacuum condense star) is described in [5–10]**. There are many investigations of this cosmic expansion and nature of DE based on different ways. These efforts can be classified as follows: (i) **to modify the entire cosmic energy by including new components of DE** and (ii) to modify Einstein-Hilbert action to get different types of modified theories of gravity such as $f(R)$ gravity [11, 12], $R$ being the Ricci scalar; $f(T)$ gravity [13], $T$ being the torsion; $f(R,T)$ gravity [14], Gauss-Bonnet gravity, i.e., $f(G)$ modified gravity [15], etc. Here we assume only $f(T)$ gravity theory.

Since general relativity is similar to $f(T)$, this theory could be a substitute form of the generalized general relativity, named as $f(T)$ theory of gravity. The teleparallel equivalence of gravity (TEGR) gives the concept of this theory. There is defined Riemann-Cartan space-time together with Weitzenbock connections rather than Levi-Civita connections in $f(T)$ theory. **Here**, nonzero torsion and zero curvature appear in the background space-time. Einstein gives this definition of space-time to give an idea of gravitation related to tetrad and torsion. Instead of metric field, tetrad field takes an important role in dynamic field in TEGR.

**In $f(T)$ gravity, equations of motions are second-order differential equations like GR whereas equations of motion are fourth-order in $f(R)$ gravity.** So, the former one is more convenient than the latter one. Recently, a wide interest has been seen to study the $f(T)$ gravity [16–20]. There is no doubt of excellence of $f(T)$ theory to explain the cosmic acceleration and analysis on large scale (clustering of galaxies) [21]. But GR must be a fantabulous agreement with solar system test and pulsar observation [22].

In theoretical astrophysics, $f(T)$ version of BTZ black hole solutions has been calculated as $f(T)$ theory was supported for examining the effects of $f(T)$ models in 3 dimensions [23]. **Later on [24], violation of Lorentz invariance made the first violation of black hole thermodynamics** in $f(T)$ gravity. Recently there are some static solutions which are spherically symmetric with charged source in $f(T)$ theory [25]. The physical conditions have been studied [26] for the existence of astrophysical stars in $f(T)$ theory after obtaining a large group of static perfect fluid solutions [27]. Capozziello et. al [28] have shown that, instead of $f(R)$ gravity, $f(T)$ removes the singularities for the exact black hole solution in D-Dimensions. Wormhole solution has been studied under $f(T)$ gravity by Sharif and Rani [29]. They have also investigated $f(T)$ gravity for static wormhole solution to verify energy conditions [30]. Again, for charged noncommutative wormhole solutions in f(T) gravity, Sharif and Rani [31, 32] have seen that this solution exists by violating energy conditions.

Generally, perfect fluid (isotropic fluid) inside the stellar object to study stellar structure and evolution is assumed because there exists isotropic pressure inside the fluid sphere. However, present observation shows that the fluid pressure of the highly compact astrophysical objects like X-ray pulsar, Her-X-1, X-ray buster 4U 1820-30, millisecond pulsar SAXJ1804.4-3658, etc. becomes anisotropy in nature which means the pressure can be rotten into two components such that one is radial pressure ($p_r$) and the other is transverse pressure ($p_t$). Now, $\Delta = p_t - p_r$ is known as the anisotropic factor. The anisotropy may arise for the different cases such as the existence of solid core, in presence of type P superfluid, phase transition, rotation, magnetic field, mixture of two fluids, and existence of external field. Generally, strange quark matter contains u, d, and s quarks. There are two ways to classify the formation of strange matter [33]. One way is the transformation of the quark hadron phase in the early universe and the other way is the reformation of neutron stars to strange matter at ultrahigh densities. A strange star is composed of the strange matter. Again the strange star can be classified into two types: Type I strange star with $M/R > 0.3$ and Type II strange star with $0.2 < M/R < 0.3$. Depending on mass, radius, and energy density, the strange star is distinguished from the neutron star [34]. It has been the most interesting topic to study the models of anisotropic stars for the last periods in GR and modified theories of gravity [35]. There have been many discussions about anisotropic star models in [36–41]. It is becoming a scientific tool to discuss the compact star models with Krori-Barua metric [42–44]. It has been seen in [45] that neutron star solution in $f(T)$ gravity model is possible if $f(T)$ is a linear function of scalar torsion.

Recently, Abbas and his collaborations [46–50] have discussed the anisotropic compact star models in GR, $f(R)$, $f(G)$, and $f(T)$ theories in diagonal tetrad case with Krori and Barua (KB) metric. Abbas et al. [49] have studied anisotropic strange star which corresponds to quintessence dark energy model with the help equation of state $p = \alpha\rho$, where $0 < \alpha < 1$. A study of strange star with MIT bag model in the framework of $f(T)$ gravity has been done by Abbas et al. [51]. Here, our main motivation of this paper is to study the anisotropic strange star models in the framework of $f(T)$ gravity with diagonal tetrad in presence of electric field and modified Chaplygin gas. In Section 2, we give a brief idea of $f(T)$ gravity. In Section 3, we study anisotropic quintessence strange star in $f(T)$ gravity with the help of modified Chaplygin gas. In Section 4, we analyze many physical phenomenon of this whole system. By matching of two metrics, the unknown constants are found out. We also make stability analysis. In Section 5, we calculate the mass function, compactness, and surface redshift function from our model to compare with observational data and finally, in Section 6, we give the summarization.

## 2. $f(T)$ Gravity: Fundamentals

In this section, we briefly overview the basics of $f(T)$ gravity. We define the torsion and the con-torsion tensor as follows [51]:

$$T^{\alpha}_{\mu\nu} = \Gamma^{\alpha}_{\nu\mu} - \Gamma^{\alpha}_{\mu\nu} = e^{\alpha}_i \left(\partial_\mu e^i_\nu - \partial_\nu e^i_\mu\right) \quad (1)$$

$$K^{\mu\nu}_{\alpha} = -\frac{1}{2}\left(T^{\mu\nu}_{\alpha} - T^{\nu\mu}_{\alpha} - T^{\mu\nu}_{\alpha}\right) \quad (2)$$

and the components of the tensor $S^{\mu\nu}_{\alpha}$ are defined as

$$S^{\mu\nu}_{\alpha} = \frac{1}{2}\left(K^{\mu\nu}_{\alpha} + \delta^{\mu}_{\alpha}T^{\beta\nu}_{\beta} - \delta^{\nu}_{\alpha}T^{\beta\mu}_{\beta}\right); \quad (3)$$

one can write the torsion scalar as

$$T = T^{\alpha}_{\mu\nu}S^{\mu\nu}_{\alpha} \quad (4)$$

Now, one can define the modified teleparallel action by replacing $T$ with a function of $T$, in analogy to $f(R)$ gravity [52, 53], as follows:

$$S = \int d^4x e \left[\frac{1}{16\pi}f(T) + L_{Matter}(\Phi_A)\right] \quad (5)$$

where we used $G = c = 1$ and $\Phi_A$ is matter fields.

The ordinary matter is an anisotropic fluid so that the energy-momentum tensor is given by

$$T^{\nu}_{\mu} = (\rho + p_t)u_\mu u^\nu - p_t \delta^{\nu}_{\mu} + (p_r - p_t)v_\mu v^\nu \quad (6)$$

where $u^\mu$ is the four-velocity, $v^\mu$ is radial four vectors, $\rho$ is the energy density, $p_r$ is the radial pressure, and $p_t$ is transverse pressure. Further, the energy-momentum tensor for electromagnetic field is given by

$$E^{\nu}_{\mu} = \frac{1}{4\pi}\left(g^{\delta\omega}F_{\mu\delta}F^{\nu}_{\omega} - \frac{1}{4}g^{\nu}_{\mu}F_{\delta\omega}F^{\delta\omega}\right) \quad (7)$$

where $F_{\mu\nu}$ is the Maxwell field tensor defined as

$$F_{\mu\nu} = \Phi_{\nu,\mu} - \Phi_{\mu,\nu} \quad (8)$$

and $\Phi_\mu$ is the four potential.

## 3. Anisotropic Strange Quintessence Star in $f(T)$ Gravity

We consider the KB metric [42] describing the interior spacetime of a strange star

$$ds^2 = -e^{a(r)}dt^2 + e^{b(r)}dr^2 + r^2\left(d\theta^2 + \sin^2\theta d\phi^2\right) \quad (9)$$

where we assume $a(r)$ and $b(r)$ are

$$a(r) = Br^2 + Cr^3,$$
$$b(r) = Ar^3 \tag{10}$$

where $A$, $B$, and $C$ are arbitrary constants. For the charged fluid source with density $\rho(r)$, radial pressure $p_r(r)$, and tangential pressure $p_t(r)$, the Einstein-Maxwell (EM) equations reduce to the form ($G = c = 1$) [51]

$$T(r) = \frac{2e^{-b}}{r}\left(a' + \frac{1}{r}\right) \tag{11}$$

$$T'(r) = \frac{2e^{-b}}{r}\left(a'' + \frac{1}{r^2} - T\left(b' + \frac{1}{r}\right)\right) \tag{12}$$

where the prime ' denotes the derivative with respect to the radial coordinate $r$.

Now the equations of motion for anisotropic fluid are [51]

$$4\pi\rho + E^2 = \frac{f}{4} - \left(T - \frac{1}{r^2} - \frac{e^{-b}}{r}(a'+b')\right)\frac{f_T}{2} \tag{13}$$

$$4\pi p_r - E^2 = \left(T - \frac{1}{r^2}\right)\frac{f_T}{2} - \frac{f}{4} \tag{14}$$

$$4\pi p_t + E^2 = \left[\frac{T}{2} + e^{-b}\left(\frac{a''}{2} + \left(\frac{a'}{4} + \frac{1}{2r}\right)\right)\right]\frac{f_T}{2} - \frac{f}{4} \tag{15}$$

$$\frac{\cot\theta}{2r^2}T' f_{TT} = 0 \tag{16}$$

$$E(r) = \frac{1}{r}\int_0^r 4\pi r^2 \sigma e^{\lambda/2} dr = \frac{q(r)}{r^2} \tag{17}$$

where $q(r)$ is the total charge within a sphere of radius $r$.

We introduce the modified Chaplygin gas (MCG) having equation of state [54]

$$p_r = \xi\rho - \frac{\zeta}{\rho^\alpha} \tag{18}$$

where $\xi$, $\alpha$, and $\zeta$ are free parameters of the model.

From (16) we get

$$f(T) = \beta T + \beta_1 \tag{19}$$

where $\beta$ and $\beta_1$ are integration constants and we assume $\beta_1 = 0$ for simple case.

Now from (10), (11), (13), (14), (18), and (19) we obtain the equation in $\rho$

$$8\pi(1+\xi)\rho^{\alpha+1} - \beta e^{-Ar^3}(2B + 3Cr + 3Ar)\rho^\alpha - 8\pi\zeta = 0 \tag{20}$$

Here we take $\alpha = 1$; then (20) reduces to the quadratic equation in $\rho$

$$8\pi(1+\xi)\rho^2 - \beta e^{-Ar^3}(2B + 3Cr + 3Ar)\rho - 8\pi\zeta = 0 \tag{21}$$

Solving this equation we get the value of energy density as

$$\rho = \frac{(2B\beta + 3C\beta r + 3A\beta r) + \sqrt{256\zeta\pi^2 e^{2Ar^3}(1+\xi) + (2B\beta + 3C\beta r + 3A\beta r)^2}}{16\pi e^{Ar^3}(1+\xi)} \tag{22}$$

and corresponding components are

$$p_r = \frac{\xi(2B\beta + 3C\beta r + 3A\beta r) + \xi\sqrt{256\zeta\pi^2 e^{2Ar^3}(1+\xi) + (2B\beta + 3C\beta r + 3A\beta r)^2}}{16\pi e^{Ar^3}(1+\xi)}$$
$$- \frac{\zeta}{\left((2B\beta + 3C\beta r + 3A\beta r) + \sqrt{256\zeta\pi^2 e^{2Ar^3}(1+\xi) + (2B\beta + 3C\beta r + 3A\beta r)^2}\right)/16\pi e^{Ar^3}(1+\xi)} \tag{23}$$

$$\rho + 3p_r = \frac{(1+3\xi)(2B\beta + 3C\beta r + 3A\beta r) + (1+3\xi)\sqrt{256\zeta\pi^2 e^{2Ar^3}(1+\xi) + (2B\beta + 3C\beta r + 3A\beta r)^2}}{16\pi e^{Ar^3}(1+\xi)}$$
$$- \frac{3\zeta}{\left((2B\beta + 3C\beta r + 3A\beta r) + \sqrt{256\zeta\pi^2 e^{2Ar^3}(1+\xi) + (2B\beta + 3C\beta r + 3A\beta r)^2}\right)/16\pi e^{Ar^3}(1+\xi)} \tag{24}$$

$$E^2 = \frac{\beta}{2r^2}e^{Ar^3}\left(-1 + 3Ar^3\right) + \frac{\beta}{2r^2} - \frac{(2B\beta + 3C\beta r + 3A\beta r) + \sqrt{256\zeta\pi^2 e^{2Ar^3}(1+\xi) + (2B\beta + 3C\beta r + 3A\beta r)^2}}{4e^{Ar^3}(1+\xi)} \tag{25}$$

$$p_t = \frac{\beta e^{-Ar^3}}{8\pi}\left\{2B + 3C - 3Ar + \frac{3}{2}Br^3(2C+A) + \frac{3}{4}r(C-A)(3Cr^3+2) + \frac{1}{r^2}\right\} - \frac{\beta}{8\pi r^2}$$
$$+ \frac{(2B\beta + 3C\beta r + 3A\beta r) + \sqrt{256\zeta\pi^2 e^{2Ar^3}(1+\xi) + (2B\beta + 3C\beta r + 3A\beta r)^2}}{16\pi e^{Ar^3}(1+\xi)}$$
(26)

Now from Figures 1 and 2, we conclude that anisotropic strange star in $f(T)$ gravity with modified Chaplygin gas acts as a dark energy candidate due to $\rho > 0$, $p_r < 0$. Again with the help of Figures 3 and 4, we notice that the equation of state $w = p_r/\rho$ lies between $-1/3$ and $-1$; i.e., the corresponding model belongs to quintessence phase not phantom phase.

The amount of net charge inside a sphere having radius r is

$$q = r^2\sqrt{\frac{\beta}{2r^2}e^{Ar^3}(-1+3Ar^3) + \frac{\beta}{2r^2} - \frac{(2B\beta + 3C\beta r + 3A\beta r) + \sqrt{256\zeta\pi^2 e^{2Ar^3}(1+\xi) + (2B\beta + 3C\beta r + 3A\beta r)^2}}{4e^{Ar^3}(1+\xi)}}$$
$$-\frac{\zeta}{\left(2B\beta + \sqrt{256\zeta\pi^2(1+\xi) + 4B^2\beta^2}\right)/16\pi(1+\xi)}$$
(27)

(29)

## 4. Physical Analysis

The central density $\rho_0$ and central radial pressure $p_0$ are given by

$$\rho_0 = \rho(r=0) = \frac{2B\beta + \sqrt{256\zeta\pi^2(1+\xi) + 4B^2\beta^2}}{16\pi(1+\xi)} \quad (28)$$

and

$$p_0 = p_r(r=0) = \frac{2B\beta\xi + \xi\sqrt{256\zeta\pi^2(1+\xi) + 4B^2\beta^2}}{16\pi(1+\xi)}$$

In this section, we investigate the nature of the anisotropic compact star as the following subsection.

Figures 1–5 represent the plots by taking $B = 5$, $C = 1$, $A = 2$, $\xi = 2$, and $\zeta = 1$.

*4.1. Anisotropic Behavior.* Now we take the derivatives of (22) and (23) with respect to $r$, given by

$$\frac{d\rho}{dr} = \frac{3\beta(C+A) + \left(768\zeta e^{2Ar^3}A\pi^2(1+\xi)r^2 + 768A\zeta e^{2Ar^3}\pi^2(1+\xi)r^2 + 2(3C\beta + 3A\beta)(2B\beta + 3Cr\beta + 3Ar\beta)\right)/2\sqrt{256\zeta e^{2Ar^3}\pi^2(1+\xi) + (2B\beta + 3Cr\beta + 3Ar\beta)^2}}{16e^{Ar^3}\pi(1+\xi)}$$
$$- \frac{24e^{Ar^3}A\pi r^2(1+\xi)\left\{\beta(2B + 3Cr + 3Ar) + \sqrt{256\zeta e^{2Ar^3}\pi^2(1+\xi) + (2B\beta + 3Cr\beta + 3Ar\beta)^2}\right\}}{16e^{Ar^3}\pi(1+\xi)^2}$$
(30)

and

$$\frac{dp_r}{dr} = \frac{\xi\left\{3\beta(C+A) + \left(1536\zeta e^{2Ar^3}A\pi^2(1+\xi)r^2 + 2(3C\beta + 3A\beta)(2B\beta + 3Cr\beta + 3Ar\beta)\right)/2\sqrt{256\zeta e^{2Ar^3}\pi^2(1+\xi) + (2B\beta + 3Cr\beta + 3Ar\beta)^2}\right\}}{16e^{Ar^3}\pi(1+\xi)}$$
$$+ \frac{16\zeta e^{Ar^3}\pi(1+\xi) + \left\{3\beta(C+A) + \left(1536\zeta e^{2Ar^3}A\pi^2(1+\xi)r^2 + 2(3C\beta + 3A\beta)(2B\beta + 3Cr\beta + 3Ar\beta)\right)/2\sqrt{256\zeta e^{2Ar^3}\pi^2(1+\xi) + (2B\beta + 3Cr\beta + 3Ar\beta)^2}\right\}}{\left\{\beta(2B + 3Cr + 3Ar) + \sqrt{256\zeta e^{2Ar^3}\pi^2(1+\xi) + (2B\beta + 3Cr\beta + 3Ar\beta)^2}\right\}^2}$$
$$- \frac{48\zeta e^{Ar^3}A\pi r^2(1+\xi)}{\beta(2B + 3Cr + 3Ar) + \sqrt{256\zeta e^{2Ar^3}\pi^2(1+\xi) + (2B\beta + 3Cr\beta + 3Ar\beta)^2}}$$
$$- \frac{24\xi e^{Ar^3}A\pi r^2(1+\xi)\left\{\beta(2B + 3Cr + 3Ar) + \sqrt{256\zeta e^{2Ar^3}\pi^2(1+\xi) + (2B\beta + 3Cr\beta + 3Ar\beta)^2}\right\}}{128e^{2Ar^3}\pi^2(1+\xi)^2}$$
(31)

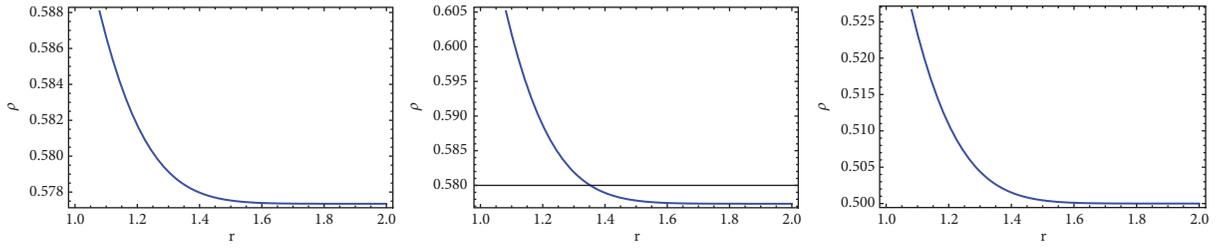

Figure 1: This figure represents the variation of $\rho$ versus $r$ (km) for the strange star taking $\beta = 1$, $\beta = 2$, and $\beta = 3$.

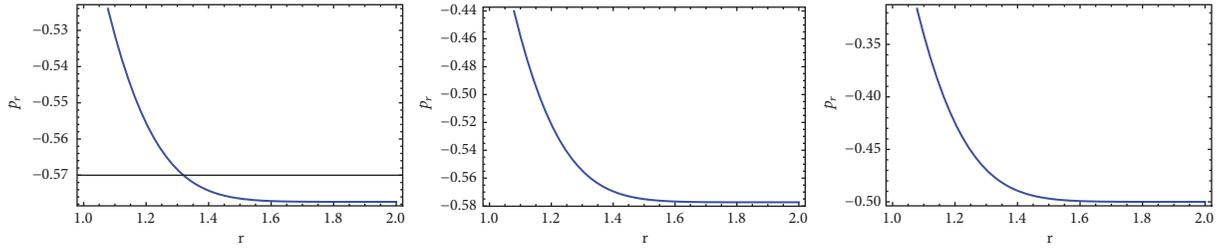

Figure 2: This figure represents the variation of $p_r$ versus $r$ (km) for the strange star taking $\beta = 1$, $\beta = 2$, and $\beta = 3$.

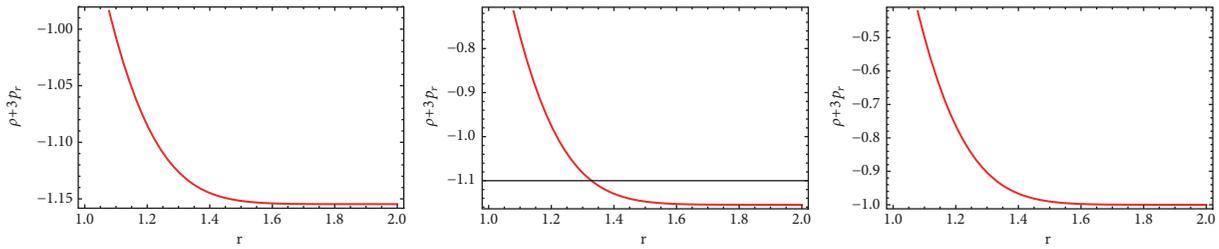

Figure 3: This figure represents the variation of $\rho + 3p_r$ versus $r$ (km) for the strange star taking $\beta = 1$, $\beta = 2$, and $\beta = 3$.

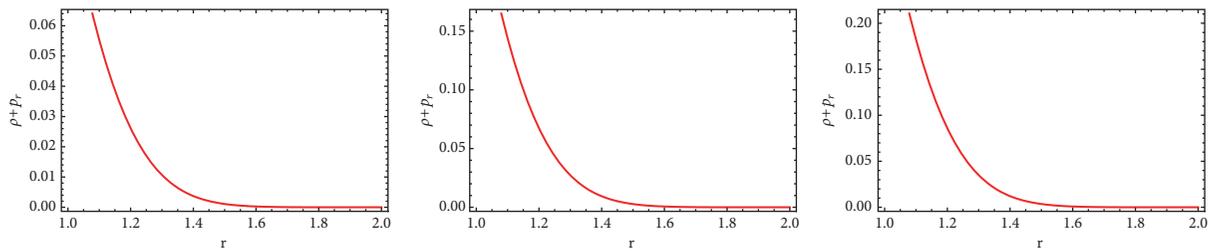

Figure 4: This figure represents the variation of $\rho + p_r$ versus $r$ (km) for the strange star taking $\beta = 1$, $\beta = 2$, and $\beta = 3$.

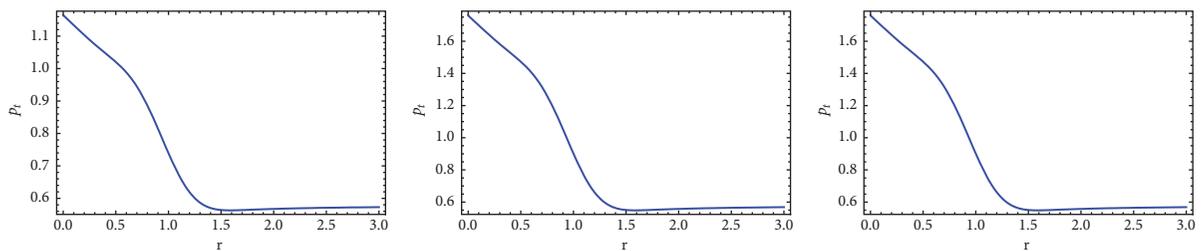

Figure 5: This figure represents the variation of $p_t$ versus $r$ (km) for the strange star taking $\beta = 1$, $\beta = 2$, and $\beta = 3$.

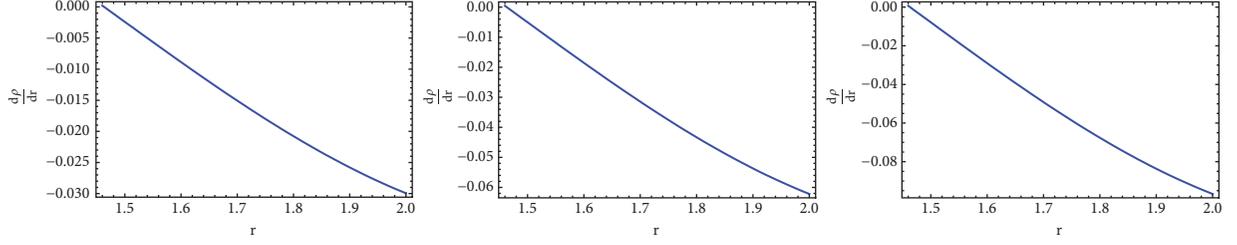

FIGURE 6: This figure represents the variation of $d\rho/dr$ versus $r$ (km) for the strange star taking $\beta = 1$, $\beta = 2$, and $\beta = 3$.

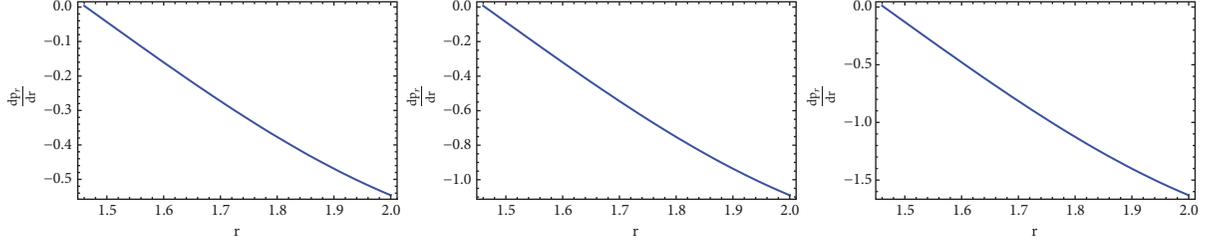

FIGURE 7: This figure represents the variation of $dp_r/dr$ versus $r$ (km) for strange the star taking $\beta = 1$, $\beta = 2$, and $\beta = 3$.

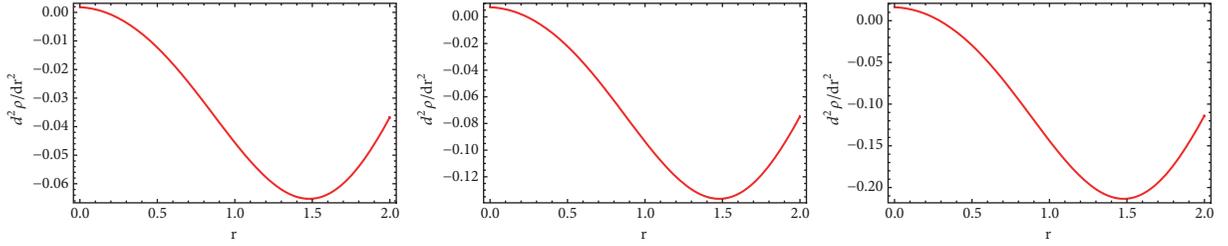

FIGURE 8: This figure represents the variation of $d^2\rho/dr^2$ versus $r$ (km) for the strange star taking $\beta = 1$, $\beta = 2$, and $\beta = 3$.

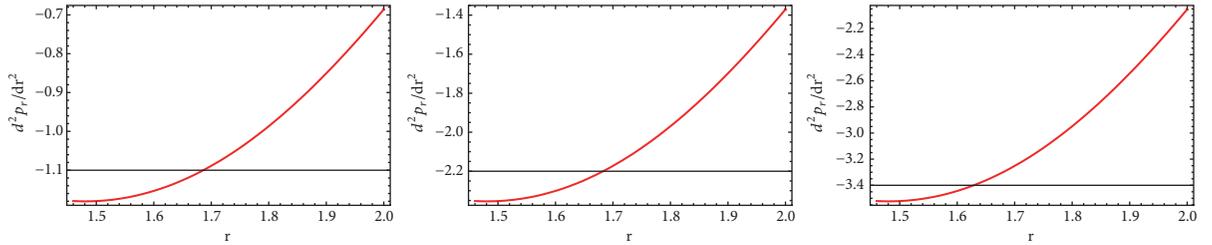

FIGURE 9: This figure represents the variation of $d^2p_r/dr^2$ versus $r$ (km) for the strange star taking $\beta = 1$, $\beta = 2$, and $\beta = 3$.

Now we present the evolution of $d\rho/dr$ and $dp_r/dr$ by Figures 6 and 7. Figure 6 shows that $d\rho/dr$ decreases keeping $d\rho/dr < 0$ (as energy density decreases) and Figure 7 shows that $dp_r/dr$ decreases keeping $dp_r/dr < 0$ (as for dark energy pressure is negatively very high; i.e., pressure decreases negatively). From Figures 6, 7, 8, and 9 we notice that, at $r = 1.46$,

$$\frac{d\rho}{dr} = 0,$$
$$\frac{dp_r}{dr} = 0,$$
$$\frac{d^2\rho}{dr^2} < 0,$$
$$\frac{d^2p_r}{dr^2} < 0. \tag{32}$$

This points out that the energy density and radial pressure have maximum value at $r = 1.46$ of the quintessence strange star model in $f(T)$ gravity.

Now the anisotropic stress ($\Delta = p_t - p_r$) is as follows [51]:

$$\Delta = \frac{\beta e^{-Ar^3}}{8\pi} \left\{ 2B + 3C - 3Ar + \frac{3}{2}Br^3 (2C + A) + (C - A)\left(\frac{9}{4}Cr^4 + \frac{3}{2}r\right) + \frac{1}{r^2} \right\} - \frac{\beta}{8\pi r^2}$$
$$+ \frac{(2B\beta + 3C\beta r + 3A\beta r) + \sqrt{256\zeta\pi^2 e^{2Ar^3}(1+\xi) + (2B\beta + 3C\beta r + 3A\beta r)^2}}{16\pi e^{Ar^3}(1+\xi)}$$
$$- \frac{\xi(2B\beta + 3C\beta r + 3A\beta r) + \xi\sqrt{256\zeta\pi^2 e^{2Ar^3}(1+\xi) + (2B\beta + 3C\beta r + 3A\beta r)^2}}{16\pi e^{Ar^3}(1+\xi)}$$
$$- \frac{\zeta}{\left((2B\beta + 3C\beta r + 3A\beta r) + \sqrt{256\zeta\pi^2 e^{2Ar^3}(1+\xi) + (2B\beta + 3C\beta r + 3A\beta r)^2}\right)/16\pi e^{Ar^3}(1+\xi)} \tag{33}$$

From Figure 10, we notice that $\Delta > 0$ for $\beta = 1, -6$ which imply that the anisotropic stress is outwardly directed and there exists repulsive gravitational force for the strange star and for $\beta = -15$, $\Delta < 0$ in somewhere implying the existence of attractive gravitational force and $\Delta > 0$ in the remaining part implying the existence of repulsive gravitational force of the strange star.

Figure 11 shows that $E^2$ is decreasing with the increment of the radial coordinate.

*4.2. Energy Conditions.* Energy conditions are very useful tools to discuss cosmological geometry in general relativity and modified gravity [10, 48, 51]. These include null energy condition (NEC), weak energy condition (WEC), and strong energy condition (SEC), given as

$$\text{NEC: } \rho + \frac{E^2}{8\pi} \geq 0,$$
$$\text{WEC: } \rho + p_r \geq 0,$$
$$\rho + p_t + \frac{E^2}{4\pi} \geq 0, \tag{34}$$
$$\text{SEC: } \rho + p_r + 2p_t + \frac{E^2}{4\pi} \geq 0.$$

Figures 6, 7, 8, and 9 represent the plots of $d\rho/dr$, $dp_r/dr$, $d^2\rho/dr^2$, and $d^2p_r/dr^2$ with respect to $r$ to show the maximality of density and radial pressure at $r = 1.46$ of the strange star taking $B = 1$, $C = 7$, $A = 0.1$, $\xi = 9$, and $\zeta = 10$.

Figure 10 represents the plots of $\Delta$ with respect to $r$ to show the presence of repulsive and attractive force of the strange star and Figure 11 represents the plot of $E^2$ with respect to $r$ taking $B = 10$, $C = 1$, $A = 10$, $\xi = 2$, and $\zeta = 1$.

From Figures 12, 4, 13, and 14, we observe that the interior of our proposed strange star model satisfies all energy conditions.

*4.3. Matching Conditions.* Many authors have worked on the matching condition to compare the exterior solution with the interior solution [10, 47, 49, 51]. We correspond the exterior geometry with our interior solution, evoked by the Schwarzschild solution which is given by the line element

$$ds^2 = -\left(1 - \frac{2M}{r}\right)dt^2 + \left(1 - \frac{2M}{r}\right)^{-1}dr^2 + r^2(d\theta^2 + \sin^2\theta d\phi^2). \tag{35}$$

The continuity of the metric components $g_{tt}$, $g_{rr}$, and $\partial g_{tt}/\partial r$ at the boundary surface $r = R$ yields

$$g_{tt}^- = g_{tt}^+,$$
$$g_{rr}^- = g_{rr}^+, \tag{36}$$
$$\frac{\partial g_{tt}^-}{\partial r} = \frac{\partial g_{tt}^+}{\partial r},$$

where − and + indicate interior and exterior solutions. Now, using (36) and the metrics (9) and (35), we have

$$A = -\frac{1}{R^3}\ln\left(1 - \frac{2M}{R}\right),$$
$$B = \frac{3}{R^2}\ln\left(1 - \frac{2M}{R}\right) - \frac{2M}{R^3}\left(1 - \frac{2M}{R}\right)^{-1}, \tag{37}$$
$$C = \frac{2M}{R^4}\left(1 - \frac{2M}{R}\right)^{-1} - \frac{2}{R^3}\ln\left(1 - \frac{2M}{R}\right).$$

For the values of $M$ and $R$ for a strange stars, we compute the constants $A$, $B$, and $C$, specified as in Table 1.

*4.4. Stability.* Now we calculate the two sound speed squares $v_{sr}^2$, $v_{st}^2$ for the radial and transverse coordinate, respectively. Herrera [55] introduced cracking concept and developed a new technique to examine potential stability for the matter. If we investigate the sign of the difference $v_{st}^2 - v_{sr}^2$ then we can conclude whether our strange star is potential stable or not; i.e., if the radial speed sound is greater than the transverse speed sound, then there exists potentially stable region; otherwise, the region will be potentially unstable [10, 47, 49, 51]. **It is clear from Figures 15 and 16 that** $0 < v_{sr}^2 \leq 1$ **and** $0 < v_{st}^2 \leq 1$ **always within the stellar objects.** From Figure 17, we see that the corresponding difference is

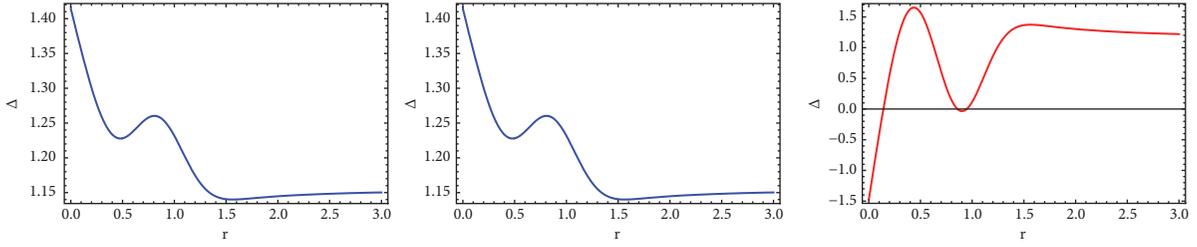

FIGURE 10: This figure represents the variation of $\Delta$ versus $r$ (km) for the strange star taking $\beta = 1$, $\beta = -6$, and $\beta = -15$.

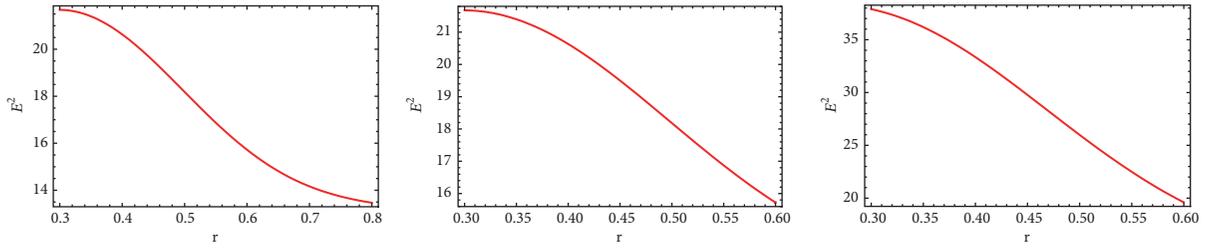

FIGURE 11: This figure represents the variation of $E^2$ versus $r$ (km) for the strange star taking $\beta = 1$, $\beta = 2$, and $\beta = 3$.

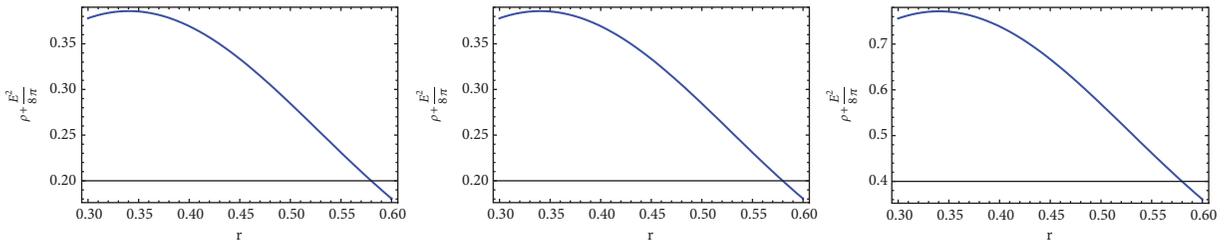

FIGURE 12: This figure represents the variation of $\rho + E^2/8\pi$ versus $r$ (km) for the strange star taking $\beta = 1$, $\beta = 2$, and $\beta = 3$.

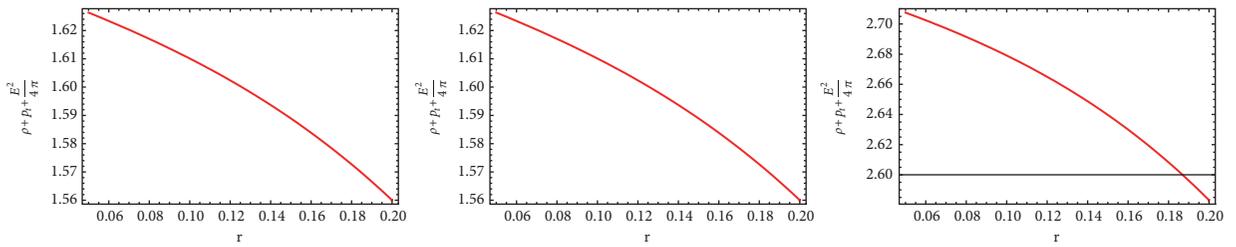

FIGURE 13: This figure represents the variation of $\rho + p_t + E^2/4\pi$ versus $r$ (km) for the strange star taking $\beta = 1$, $\beta = 2$, and $\beta = 3$.

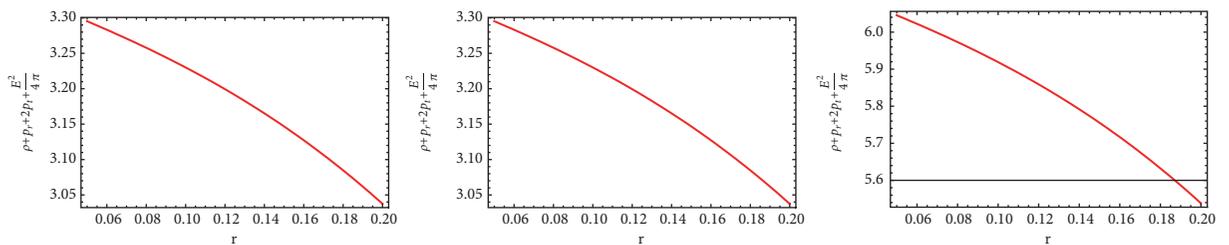

FIGURE 14: This figure represents the variation of $\rho + p_r + 2p_t + E^2/4\pi$ versus $r$ (km) for the strange star taking $\beta = 1$, $\beta = 2$, and $\beta = 3$.

TABLE 1: The values of $A$, $B$, and $C$ have been obtained using (37).

| Compact Stars | $M(M_\odot)$ | $R(Km)$ | $A(Km^{-2})$ | $B(Km^{-2})$ | $C(Km^{-2})$ |
|---|---|---|---|---|---|
| SAX J 1808.4 − 3658(SS1) | 1.435 | 7.07 | 0.001473644346 | −0.044926791 | 0.004880923098 |
| 4U 1820 − 30 | 2.25 | 10 | 0.0005978370008 | −0.026116928 | 0.00201385582 |
| Vela X − 12 | 1.77 | 9.99 | 0.0004388196046 | −0.018169038 | 0.001428126229 |
| PSR J 1614 − 2230 | 1.97 | 10.3 | 0.000441203995 | −0.019472555 | 0.00144933537 |

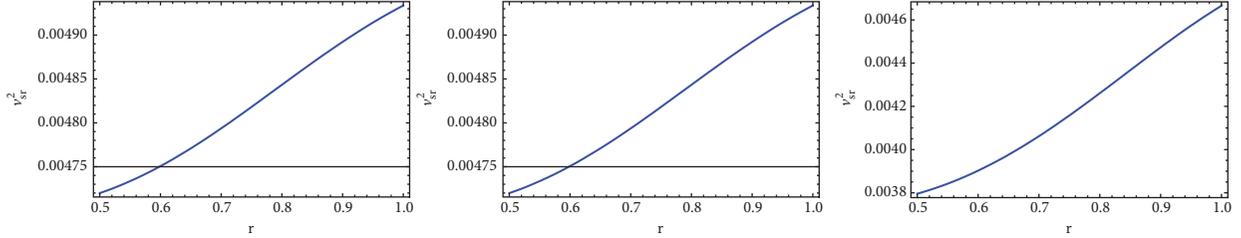

FIGURE 15: This figure represents the variation of $v_{sr}^2$ versus $r$ (km) for the strange star taking $\beta = 1$, $\beta = 2$, and $\beta = 3$.

negative which means the radial speed sound is greater than the transverse speed sound which implies that our proposed strange star model is potentially stable in the framework of $f(T)$ gravity. **Again Figure 18 shows that $|v_{st}^2 - v_{sr}^2| \le 1$ is satisfied [56].**

Figures 12, 13, and 14 represent the plots to understand the validation of the energy conditions taking $B = 10$, $C = 1$, $A = 10$, $\xi = 2$, and $\zeta = 1$.

Figures 15, 16, 17, and 18 represent the plots to show the stability of our proposed model taking $B = 5$, $C = 1$, $A = 2$, $\xi = 2$, and $\zeta = 1$.

## 5. Some Fundamental Calculations

### 5.1. Mass Function and Compactness.
The mass function within the radius $r$ is defined as [10]

$$m(r) = \int_0^r 4\pi r^2 \rho \, dr = 2\pi \int_0^r \frac{r^2 \left\{(2B\beta + 3C\beta r + 3A\beta r) + \sqrt{256\zeta \pi^2 e^{2Ar^3}(1+\xi) + (2B\beta + 3C\beta r + 3A\beta r)^2}\right\}}{8\pi e^{Ar^3}(1+\xi)} dr \quad (38)$$

From Figure 19, we have seen that at origin the mass function is regular (i.e., $m(r) \to 0$ when $r \to 0$) and monotonic increasing with respect to radius ($r$). We have also evaluated the values of mass for a few strange stars from our model to compare these values with observational data (see Table 2).

The compactness of the star is defined by $u(r)$ [10] in the form of

$$u(r) = \frac{m(r)}{r} = \frac{2\pi}{r} \int_0^r \frac{r^2 \left\{(2B\beta + 3C\beta r + 3A\beta r) + \sqrt{256\zeta \pi^2 e^{2Ar^3}(1+\xi) + (2B\beta + 3C\beta r + 3A\beta r)^2}\right\}}{8\pi e^{Ar^3}(1+\xi)} dr \quad (39)$$

We have plotted the corresponding function given by Figure 20.

### 5.2. Relation between Mass and Radius.
In this section we discuss the mass radius relation of the strange stars. From [57], twice the maximum allowable ratio of mass to the radius for an astrophysical object is always less than 8/9 ($2M/R < 8/9$) whereas the factor $M/R$ is called "compactification factor". From Table 3, we find that the calculated values corresponding to our model lie in the expected range [34]. Compactification factor for strange star always lies between 1/4 and 1/2. The calculated values of the compactification factor of the strange stars from our model are compatible with the condition (see Table 3).

### 5.3. Surface Redshift.
The redshift function can be defined as [10, 47, 49, 51]

$$z_s = \frac{1}{\sqrt{1 - 2m(r)/r}} - 1, \quad (40)$$

where $m(r)$ has been obtained from (38). According to Bohmer and Harko, the surface redshift should be $\le 5$ for

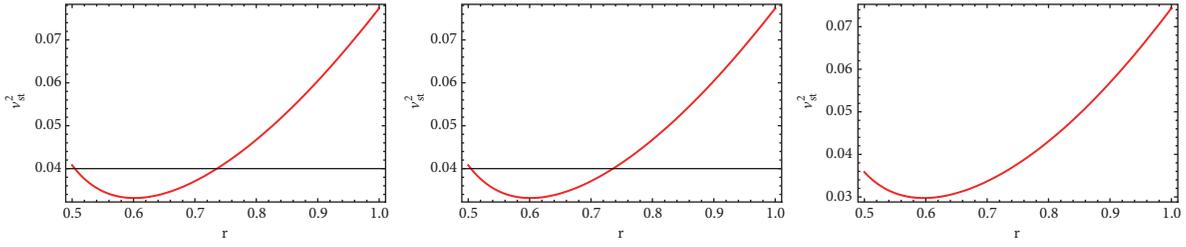

FIGURE 16: This figure represents the variation of $v_{st}^2$ versus $r$ (km) for the strange star taking $\beta = 1$, $\beta = 2$, and $\beta = 3$.

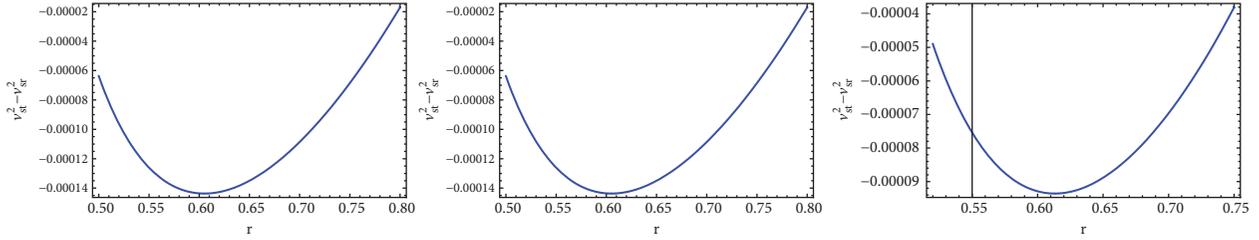

FIGURE 17: This figure represents the variation of $v_{st}^2 - v_{sr}^2$ versus $r$ (km) for the strange star taking $\beta = 1$, $\beta = 2$, and $\beta = 3$.

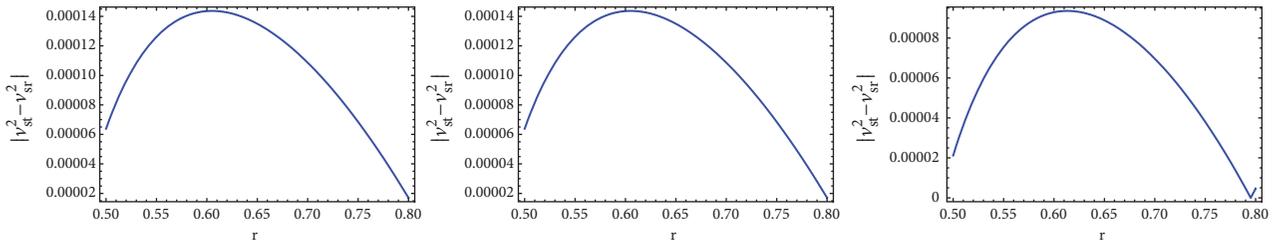

FIGURE 18: This figure represents the variation of $|v_{st}^2 - v_{sr}^2|$ versus $r$ (km) for the strange star taking $\beta = 1$, $\beta = 2$, and $\beta = 3$.

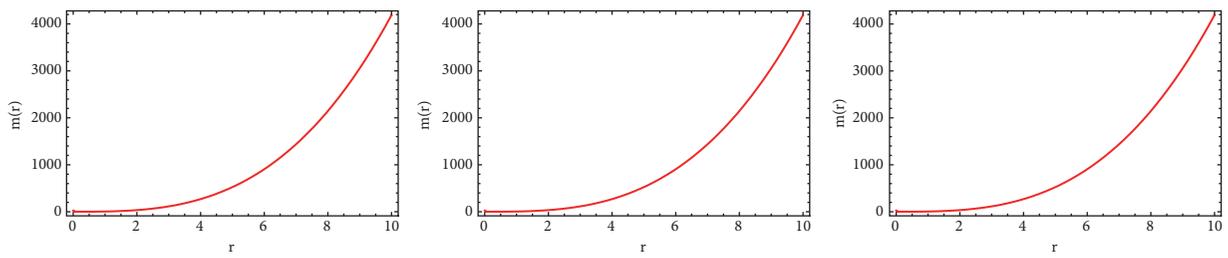

FIGURE 19: This figure represents the variation of $m(r)$ versus $r$ (km) for the strange star taking $\beta = 1$, $\beta = 2$, and $\beta = 3$.

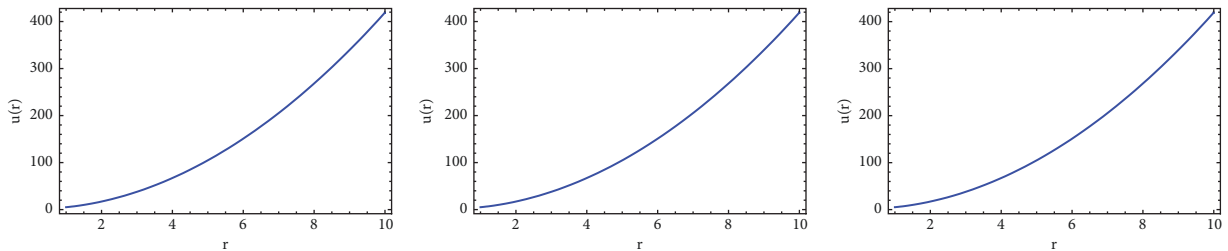

FIGURE 20: This figure represents the variation of $u(r)$ versus $r$ (km) for the strange star taking $\beta = 1$, $\beta = 2$, and $\beta = 3$.

Table 2: Calculated values of mass, energy density, and pressure from our model.

| Compact Stars | Mass standard data (in km) | Mass from model (in km) | $\rho_0 (gm/cc)$ | $\rho_R (gm/cc)$ | $p_0 (dyne/cm^2)$ |
|---|---|---|---|---|---|
| $SAX\ J\ 1808.4-3658(SS1)$ | 2.116625 | 2.0868 | $1.996428 \times 10^{-12}$ | $1.000531 \times 10^{-12}$ | $-1.001789 \times 10^{12}$ |
| $4U1820-30$ | 3.31875 | 3.34265 | $1.997923 \times 10^{-12}$ | $1.000286 \times 10^{-12}$ | $-1.001040 \times 10^{12}$ |
| $Vela\ X-12$ | 2.61075 | 2.61043 | $1.998555 \times 10^{-12}$ | $1.000253 \times 10^{-12}$ | $-1.000723 \times 10^{12}$ |
| $PSR\ J\ 1614-2230$ | 2.90575 | 2.91837 | $1.998451 \times 10^{-12}$ | $1.000239 \times 10^{-12}$ | $-1.000775 \times 10^{12}$ |

Table 3: Calculated values of the desired parameters of our model.

| Compact Stars | M/R (standard data) | M/R from model | 2M/R < 8/9 | $\rho_0/\rho_R$ | $z_s$ |
|---|---|---|---|---|---|
| $SAX\ J\ 1808.4-3658(SS1)$ | 0.299381 | 0.295163 | 0.590325 | 1.995368 | 0.562358 |
| $4U1820-30$ | 0.331875 | 0.334265 | 0.66853 | 1.997352 | 0.736912 |
| $Vela\ X-12$ | 0.266134 | 0.261304 | 0.522609 | 1.998050 | 0.447314 |
| $PSR\ J\ 1614-2230$ | 0.282112 | 0.283337 | 0.566674 | 1.997973 | 0.519122 |

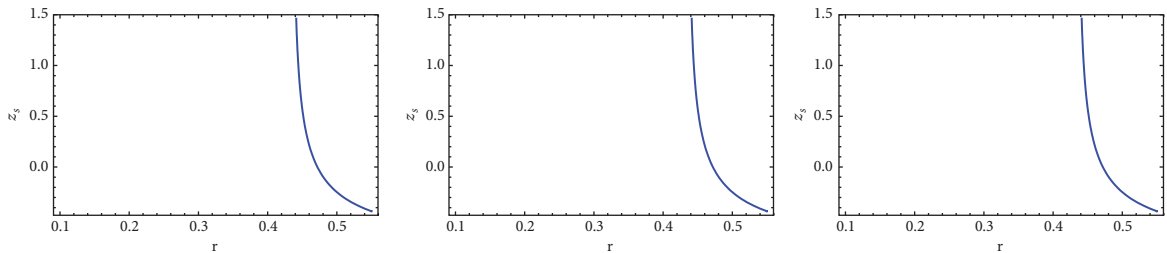

Figure 21: This figure represents the variation of $z_s$ versus $r$ (km) for the strange star taking $\beta = 0.009$, $\beta = 0.010$, and $\beta = 0.011$.

an anisotropic star in the presence of a cosmological constant [58]. We calculate the maximum surface redshift from our model in Table 3. Now, it is clear that our model for strange stars obeys the relation $z_S \leq 5$ though the cosmological constant is absent in our model which is quite reasonable.

## 6. Discussions

This paper has given out the anisotropic strange star model in $f(T)$ gravity with modified Chaplygin gas. Using the diagonal tetrad field we have obtained the equations of motion where we have solved the unknown function $f(T)$ as $\beta T + \beta_1$, $\beta$ and $\beta_1$ being constants. Then we have solved the differential equation of energy density from where we have found the value of energy density (22) of it. With the help of this energy density, we have found out radial pressure ensuring this model as a quintessence dark energy candidate from Figures 1, 2, 3, and 4. We have also noticed that both the energy density ($\rho$) and radial pressure ($p_r$) are monotonic decreasing function with respect to $r$ and they have maximum value at $r = 1.46$ by Figures 6–9. Figure 5 shows that the transverse pressure is decreasing with the rise of $r$. We have calculated anisotropic factor to see whether there exists gravitational attractive force or repulsive force for the strange star and we have studied from Figure 10 that there exists attractive force as well as repulsive gravitational force with different values of $\beta$. Here, the square of energy is monotonic decreasing with the increment of radial coordinate given by Figure 11. From Figures 12, 4, 13, and 14 we have concluded that all energy conditions are satisfied for our proposed model.

Using matching condition, the unknown parameters $A$, $B$, and $C$ have been calculated for the different strange stars from our model which is given by Table 1. By stability analysis given on the basic of Figures 15–18, we have observed that $0 < v_{sr}^2, v_{st}^2 \leq 1$, $v_{sr}^2 > v_{st}^2$, and $|v_{st}^2 - v_{sr}^2| \leq 1$ always. Finally, we have ensured that our model is potentially stable.

In Table 2, with the help of energy density (22) and radial pressure (23) we have calculated the numerical values of the mass of the different strange stars from our model to show the closeness of these values with the observational data. Also, we have obtained the values of central and surface density and central pressure for the above-mentioned strange stars from our model which have been calculated in Table 2. From Table 3, we have observed that twice the compactification factor are always less than $< 8/9$ and maximum values of the surface redshift function are always less than 5. So, our proposed model is completely rational.

Figures 19, 20, and 21 represent the plots of mass function, compactness, and surface redshift function taking the values of $A$, $B$, and $C$ from Table 1 and $\xi = 2$ and $\zeta = 1$.

## Data Availability

No data were used to support this study.

## Conflicts of Interest

The authors declare that they have no conflicts of interest.

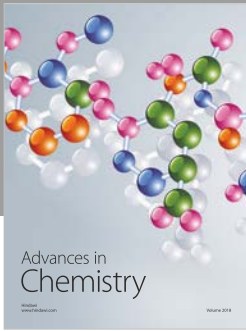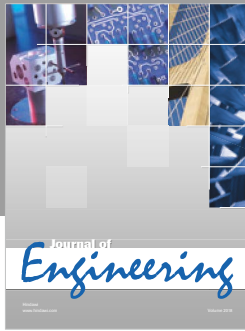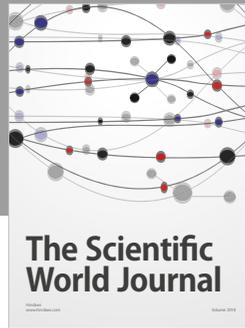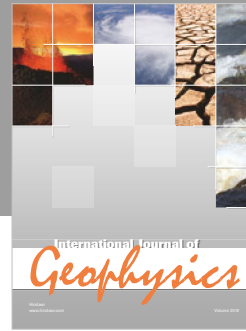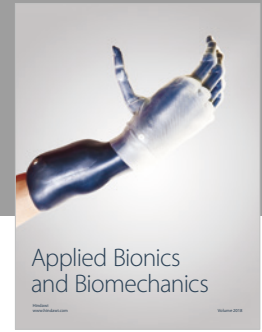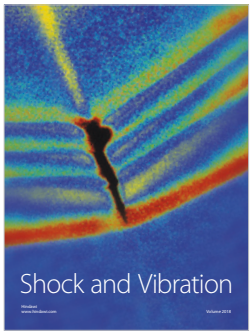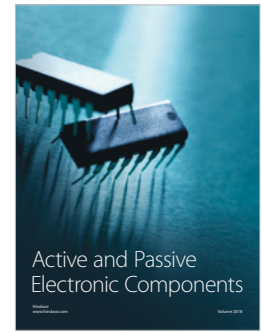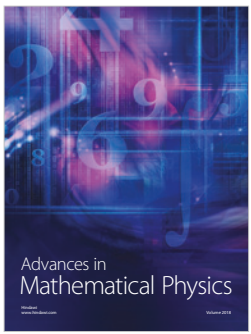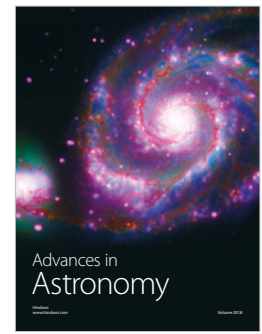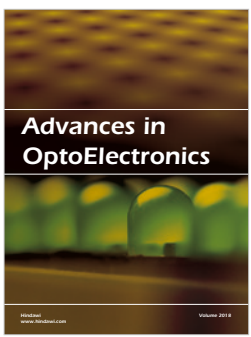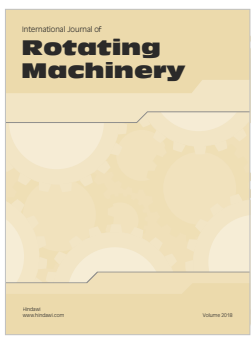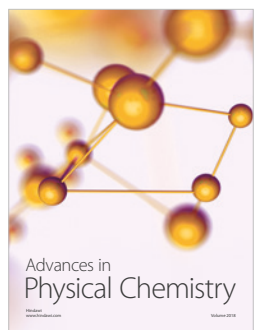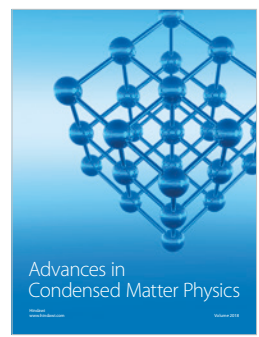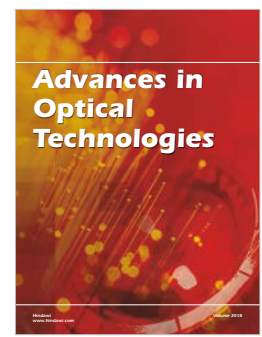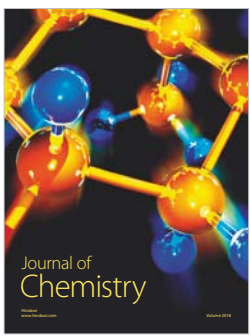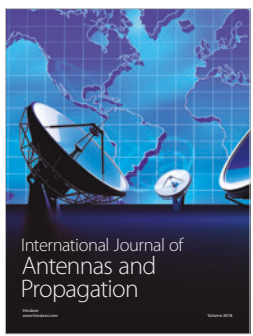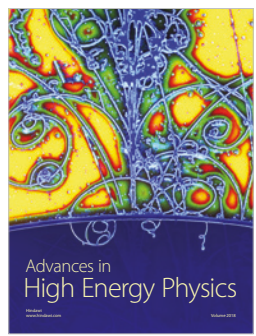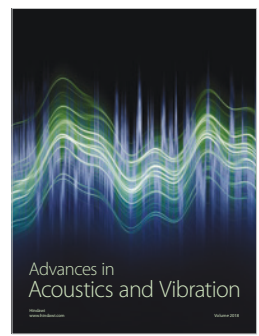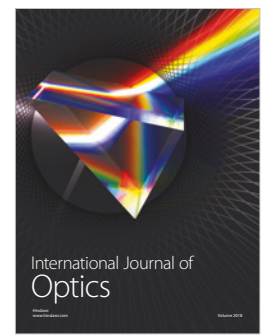